# A 40Gb/s Linear Redriver with Multi-Band Equalization in 130nm SiGe BiCMOS

Tong Liu[1], Yuanming Zhu[1], Anil Korkmaz[1], Siamak Delshadpour[2], and Samuel Palermo[1]

[1]Analog and Mixed-Signal Center, Texas A&M University, College Station, TX, USA

[2]NXP Semiconductors, Chandler, AZ, USA

liut@tamu.edu

*Abstract*—A linear redriver circuit implements multi-band equalization techniques to efficiently compensate for high-frequency channel loss and extend high-speed wireline link reach. Input and output stage emitter-follower buffers with dual AC and DC paths provide programmable low-frequency peaking for channel skin effect, while a continuous-time linear equalizer (CTLE) utilizes RC degeneration in the input stage for mid-band peaking and a subsequent feedback structure contributes to additional high-frequency peaking to compensate for the additional dielectric loss effects. A variable-gain amplifier (VGA) stage provides up to 7.1dB tunable gain and utilizes negative capacitive loads for bandwidth extension. Input and output return loss of -11.0dB and -12.2dB is respectively achieved at 20GHz with input and output T-coil stages that distribute the ESD circuitry capacitance. Fabricated in a 130nm SiGe BiCMOS process, the redriver achieves 23.5dB max peaking at 20GHz and supports a 1$V_{ppd}$ linear output swing. Per-channel power consumption is 115.2mW from a 1.8V supply.

## I. INTRODUCTION

As wireline communication data rates scale, high-frequency channel loss causes inter-symbol interference (ISI) that grows with distance. As shown in Fig. 1, high-loss channels induce significant pulse attenuation and dispersion, necessitating high-performance SERDES transceivers to employ transmit-side feed-forward equalization (FFE) and receive-side continuous-time linear equalization (CTLE) and decision-feedback equalization (DFE) to achieve the desired system bit error rate (BER). However, this results in significant transceiver complexity as the Nyquist frequency channel loss approaches 40dB. An efficient solution is to break a long channel into two moderate-loss segments with a low-complexity redriver chip in between that provides signal equalization and amplification. This redriver allows for reduced complexity in the SERDES transceivers at the ends of the overall channel and/or extended link reach.

Equalization in redrivers is typically implemented in the form of CTLE frequency peaking and output driver FFE. However, implementing FFEs in redrivers can result in high power consumption and complicate the SERDES transceivers equalization training due to the limiting amplifiers employed in the signal chain before the FFE taps breaking the typical linear channel model [1]. Also, FFE tap delay cell tuning constraints can limit multi-rate support [2]. While CTLE-only redrivers provide efficient channel loss compensation, the amount of achievable frequency

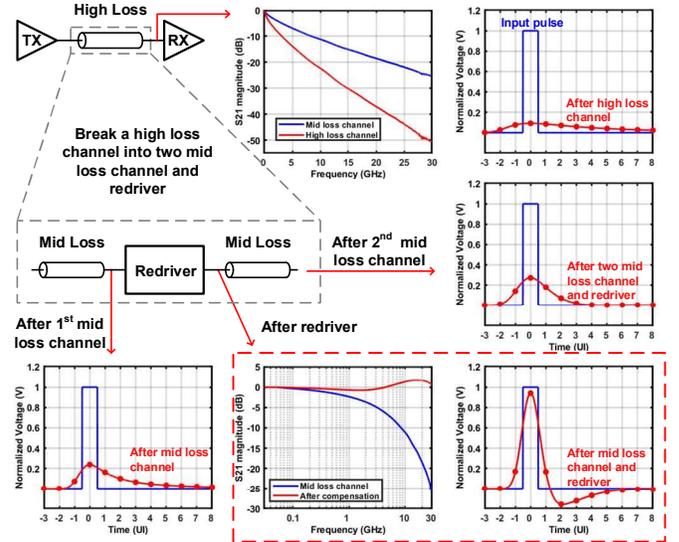

Fig. 1. High-loss channel compensation with a redriver.

peaking and ability to support a given loss profile is limited with only a single CTLE stage [3].

This work presents a linear redriver that implements multi-band equalization to provide up to 23.5dB peaking at 20GHz, which can be used ADC-based receiver front end [6][7][8][9][10]. The redriver architecture is detailed in Section II. Section III discusses the key redriver circuits, including the input and output buffers that provide low-frequency peaking, the CTLE that has RC degeneration for mid-band peaking and a subsequent feedback structure for additional high-frequency peaking, and the variable-gain amplifier (VGA) stage that employs negative capacitance for bandwidth extension. Redriver measurement results from a 130nm SiGe BiCMOS prototype are presented in Section IV. Finally, Section V concludes the paper.

## II. REDRIVER ARCHITECTURE

Fig. 2 shows the proposed linear redriver block diagram. Input T-coil structures are utilized to distribute the pad, electrostatic discharge (ESD) circuitry, and the termination and buffer capacitance in order to achieve wideband input matching. This followed by the input buffer stage that provides low-frequency peaking and sets the input common-mode of the CTLE input stage. Mid-band peaking tuning is achieved with programmable RC emitter degeneration in the CTLE input transconductance stage. A subsequent feedback structure produces a high-frequency zero-pole pair that allows for high-frequency peaking optimization. As peaking optimization can change the CTLE DC gain, a VGA with negative capacitance bandwidth extension follows to provide an overall programmable system gain. After the VGA is another buffer stage with the same topology as the input buffer. This provides additional low-frequency peaking and sets the common-mode of the current-mode logic (CML) output driver that is 50Ω terminated on chip. The output driver provides a maximum 1V$_{ppd}$ output swing onto the controlled impedance channel. Output T-coil structures are utilized to distribute the output driver, ESD, and pad capacitance to allow for wideband output matching.

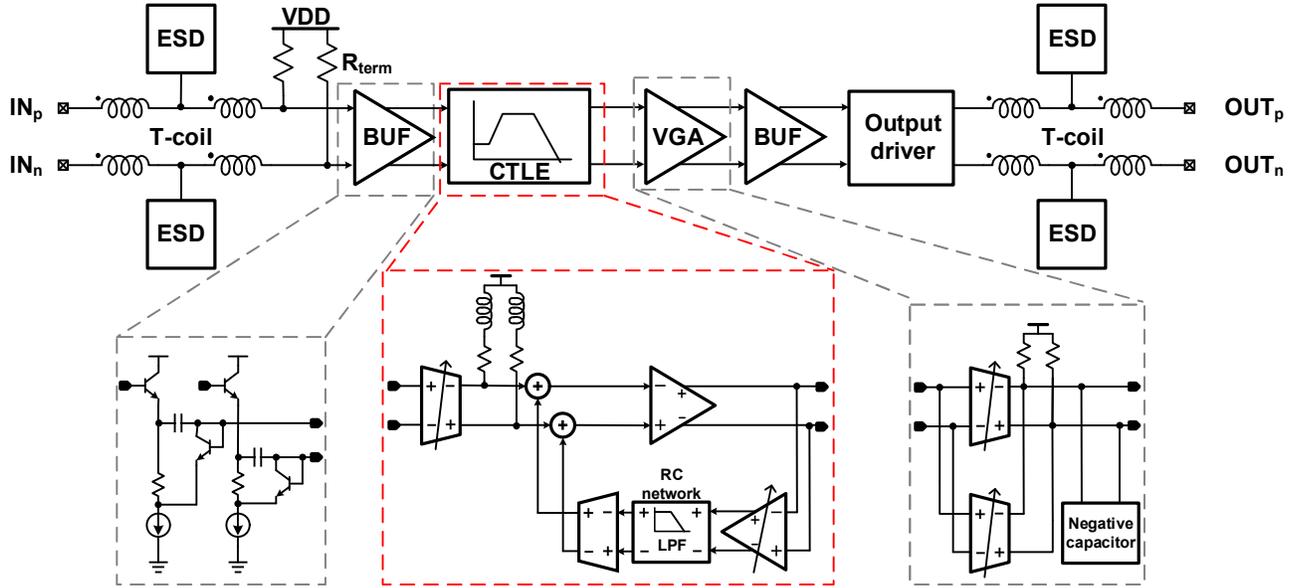

Fig. 2. Redriver block diagram.

## III. REDRIVER CIRCUIT DESIGN

### A. Input/Output Buffers

The input and output buffer circuitry is shown in Fig. 3 [3]. This emitter-follower topology utilizes dual DC and AC paths to set the desired output common-mode and provide low-frequency peaking. At high frequencies the AC-coupling capacitor shorts the input transistor's emitter node to the output with a small amount of loss. Lower-frequency signal components traverse through the $R_{buf}$ and diode-connected output transistor path and experience additional voltage-division loss. Two current DACs are utilized to tune the output common-mode that is set by the difference of $I_1$ and $I_2$.

$$V_{out,DC} = V_{in,DC} - (I_1 - I_2)R_{buf} \quad (1)$$

The $I_2$ absolute value is utilized to tune the output node impedance and position the low-frequency zero to compensate for channel skin effect, with a smaller $I_2$ resulting in a lower-frequency zero. The combined

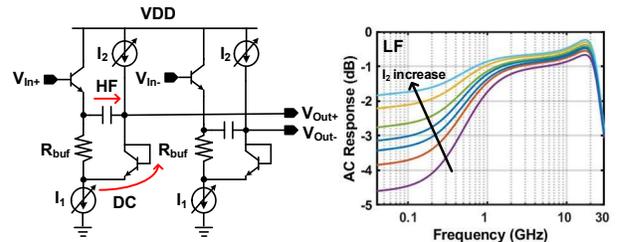

Fig. 3. Input/output buffer schematic and combined simulated low-frequency peaking response.

response of the input and output stage buffers yields a maximum of 3.5dB low-frequency peaking.

### B. Multi-Stage CTLE

Fig. 4(a) shows the multi-stage CTLE that provides mid-band and high-frequency peaking [4]. The input transconductance stage employs RC degeneration to generate a zero in the AC response.

$$f_{z1} \approx 1/(2\pi R_{ei} C_{ei}) \quad (2)$$

Increasing $R_{ei}$ moves this zero to a lower frequency and also reduces the DC gain, resulting in a larger amount of overall peaking (Fig. 4(b)). As the parasitic capacitance of the bottom current sources limits the maximum zero position, this tunable input-stage RC degeneration is utilized to optimize the mid-band peaking. This is apparent when the VGA is adjusted to provide a constant 0dB low-frequency gain for different $R_{ei}$ settings and the slope changes in the mid-band region between 2-10GHz (Fig. 4(c)).

Optimization of the high-frequency peaking is achieved with the subsequent feedback structure consisting of forward-path amplifier $A$ whose output is fed back through an amplifier stage, low-pass filter, and transconductance stage to current sum at the first-stage output load where shunt peaking is employed to extend the bandwidth. The low-pass filter pole in the feedback path generates a zero-pole pair in the closed-loop transfer function.

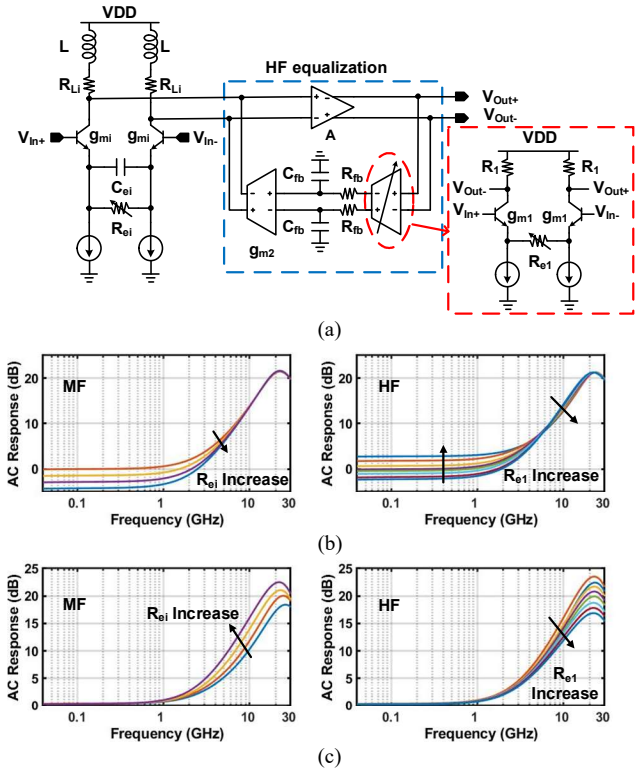

Fig. 4. Multi-stage CTLE: (a) schematic and simulated AC response with (b) fixed VGA setting and (c) variable VGA settings to achieve 0dB DC gain.

$$f_{z2} \approx 1/(2\pi R_{fb} C_{fb}) \tag{3}$$

$$f_p \approx f_{z2}(1 + g_{m1} R_1 g_{m2} R_{Li} A/(1 + g_{m1} R_{e1}/2)) \tag{4}$$

While previously this feedback structure was used for low-frequency equalization [5], the proposed design modifies this for tunable high-frequency peaking without significant loading in the main forward signal path. Increasing $R_{e1}$ moves the pole due to the feedback structure to a slightly lower frequency and also increases DC gain, resulting in a smaller amount of overall peaking (Fig. 4(b)). As evident when the VGA is adjusted to provide a constant 0dB DC gain for different $R_{e1}$ settings, tuning this feedback path primarily impacts the slope between 5-20GHz and the overall high-frequency peaking value (Fig. 4(c)).

### C. VGA

After the CTLE is the VGA stage shown in Fig. 5. This VGA employs two parallel transconductance stages with programmable emitter resistive degeneration and bias settings to allow for coarse and fine gain tuning from -3.7 to 3.4dB. As the VGA bandwidth can limit the achievable peaking, a negative capacitor cell is utilized at the output. This negative capacitance is activated for high gain settings to achieve significant bandwidth extension.

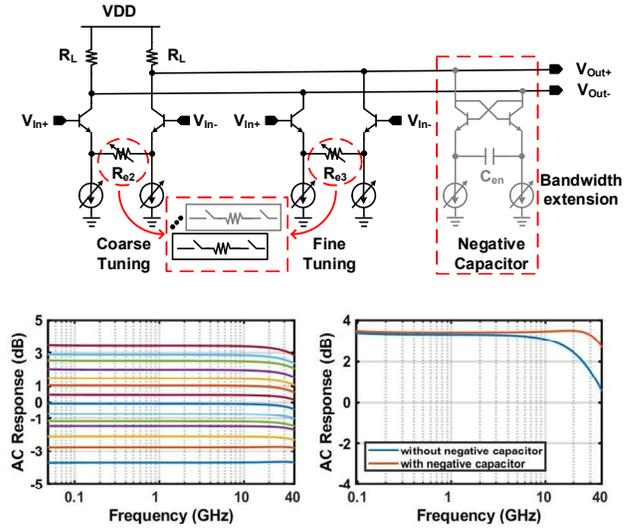

Fig. 5. VGA schematic and simulated AC response showing the gain tuning range and negative capacitor bandwidth extension.

### IV. EXPERIMENTAL RESULTS

Fig. 6(a) shows the chip micrograph of the redriver prototype with two identical channels, which was fabricated in a 130nm SiGe BiCMOS process. Total chip area is 1.22mm$^2$, with a single channel occupying 0.61mm$^2$.

Probe testing was performed to obtain the measured s-parameters shown in Fig. 7. Input and output return loss of -11.0dB and -12.2dB is respectively achieved at 20GHz with input and output T-coil stages that distribute the ESD circuitry capacitance. Configuring the VGA to achieve 0dB DC gain, the proposed redriver can achieve peaking from 16dB to 23.5dB at 20GHz.

Eye diagram and BER measurements are performing with the chip wirebonded to a PCB with a 2.2" trace and connected to a channel board via 12" cables. Redriver BER measurements were performed at 28Gb/s using an 8" FR4 trace on the channel board, resulting in 20.8dB total loss at 14GHz (Fig. 6(b)). An FPGA is utilized to generate un-equalized PRBS31 data with 1V$_{ppd}$ swing. Fig. 8 shows that passing this data through the channel results in a closed eye at the redriver input. Optimizing the redriver settings then allows for an open eye at the redriver output. BER verification is achieved by looping back the redriver output to the FPGA, with 0.2UI timing margin and 212mV voltage margin achieved at BER=10$^{-12}$. An AWG is used for 40Gb/s eye measurements over a 3" FR4 trace on the channel board, with 14.1dB total loss at 20GHz (Fig. 6(b)). As shown in Fig. 9, the completely closed eye after the channel is effectively compensated with optimal redriver settings to produce an open 40Gb/s output eye. This shows that the redriver can extend the tolerable channel loss when combined with a moderate complexity wireline receiver.

Table I summarizes the redriver performance and compares it with other published redrivers. The multi-band equalization techniques allow the proposed redriver to achieve the highest 23.5dB max peaking at the highest 20GHz frequency. Utilizing CTLE-only equalization and operating at a relatively low 1.8V supply allows for an excellent measured power efficiency of 4.11mW/Gbps at 28Gb/s and 2.88mW/Gbps at 40Gb/s operation.

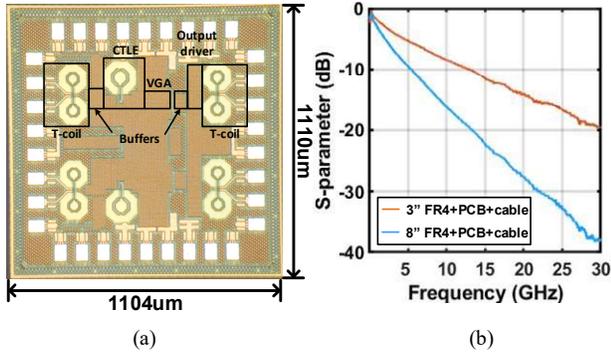

Fig. 6. (a) Chip micrograph and (b) measured channel s-parameters.

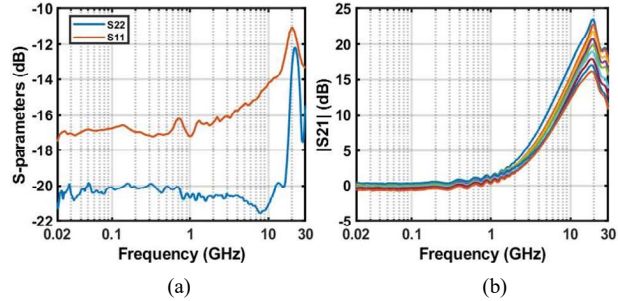

Fig. 7. Measured redriver (a) $S_{11}$, $S_{22}$ and (b) $S_{21}$ with fixed 0dB DC gain.

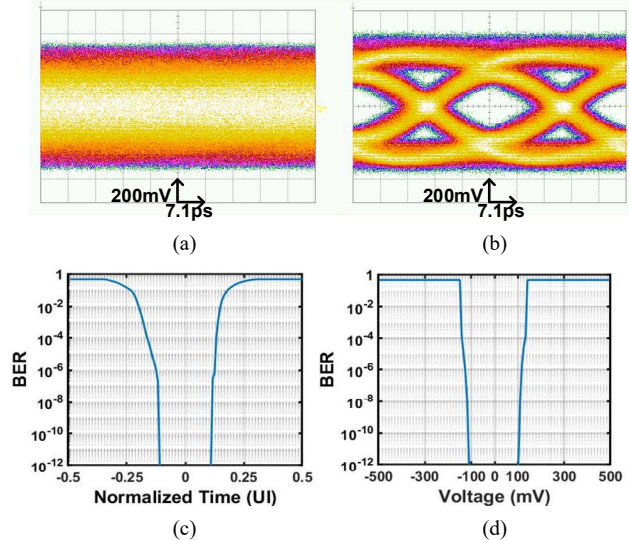

Fig. 8. Measured 28Gb/s performance: Eye diagrams (a) after test channel and (b) at redriver output. (c) Timing and (d) voltage bathtub curves.

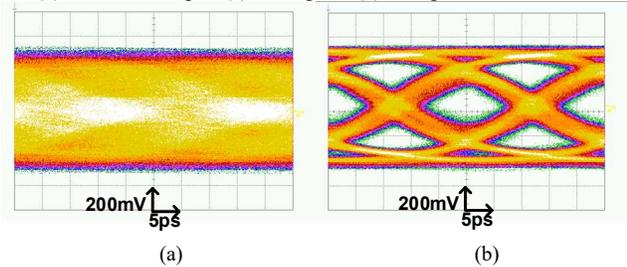

Fig. 9. Measured 40Gb/s eye diagrams (a) after test channel and (b) at redriver output.

TABLE I. PERFORMANCE SUMMARY

| | [1] | [2] | [3] | This Work |
|---|---|---|---|---|
| Technology | 130nm CMOS | 130nm BiCMOS | 250nm BiCMOS | 130nm BiCMOS |
| Data Rate (Gbps) | 5 | 25 | 20.625 | 28/40 |
| EQ Mode | FFE + CTLE | FFE + CTLE | CTLE | CTLE |
| Peak Frequency (GHz) | 2.5 | 12.5 | 10.3 | 20 |
| Max EQ Boost Gain (dB) | 12[a] | 14.4[b] | 17 | 23.5 |
| DC Gain (dB) | -14~-2 | 0 | 0 | 0 |
| Input Swing ($V_{ppd}$) | 0.1~2 | 0.5 | 0.045~1.2 | 0.167~1 |
| Output Swing ($V_{ppd}$) | 0.6~1.2 | 0.636 @25Gbps | 0.045~1.2 | 0.167~1 |
| Supply (V) | 3.3 | 3.3 | 1.8 | 1.8 |
| Power Dissipation (mW) | 272.25 | 230 | 84 | 115.2 |
| Power Efficiency (mW/Gbps) | 54.45 | 9.2 | 4.07 | 4.11 (28G) 2.88 (40G) |

[a] 6dB extra EQ boost from FFE
[b] 9.1dB extra EQ boost from FFE

V. CONCLUSION

This paper presented a linear redriver circuit that implements multi-band equalization techniques. Dual AC/DC path input and output-stage emitter follower buffers provide low-frequency peaking, while a multi-stage CTLE utilizes RC-degeneration and a feedback structure to implement mid-band and high-frequency peaking, respectively. A VGA stage with negative capacitive loads provides up to 7.1dB tunable gain with bandwidth sufficient to support 23.5dB max peaking at 20GHz. The linear-mode redriver operation makes it also applicable for PAM4 links, which is planned in future work.

ACKNOWLEDGEMENT

This work was supported by NXP Semiconductors.